# Observation of Zero Resistance in As-Electrodeposited FeSe

Aichi Yamashita[1,2,‡], Ryo Matsumoto[1,2], Masashi Tanaka[1,3], Hiroshi Hara[1,2], Kazumasa Iida[4,5], Bernhard Holzapfel[4,6], Hiroyuki Takeya[1], and Yoshihiko Takano[1,2]

[1] WPI-MANA, National Institute for Materials Science, 1-2-1 Sengen, Tsukuba, Ibaraki 305-0047, Japan

[2] Graduate School of Pure and Applied Sciences, University of Tsukuba, 1-1-4 Tennodai, Tsukuba, Ibaraki 305-8577, Japan

[3] Graduate School of Engineering, Kyushu Institute of Technology, 1-1 Sensui-cho, Tobata-ku, Kitakyushu-shi, Fukuoka 804-8550, Japan

[4] Institute for Metallic Materials, IFW Dresden, D-01171 Dresden, Germany

[5] Department of Materials Physics, Nagoya University, Nagoya 464-8603, Japan

[6] Institute for Technical Physics, Karlsruhe Institute of Technology, Hermann-von-Helmholtz-Platz 1, 76344 Eggenstein- Leopoldshafen , Germany

**Abstract**

Superconducting FeSe films were electrochemically deposited on rolling-assisted biaxially textured substrate (RABiTS) tape. We observed zero resistivity in the as-electrodeposited FeSe film prepared on the RABiTS when the appropriate voltage was applied while it was dipped into the solution. When the RABiTS tape was dipped in the solution without applying voltage, a thin Se film was deposited on the substrate. The compositional ratio of the FeSe film got closer to the stoichiometric ratio with decreasing the dipping time before applying voltage.

‡Corresponding author: Aichi Yamashita

E-mail: YAMASHITA.Aichi@nims.go.jp

Postal address: National Institute for Materials Science, 1-2-1 Sengen, Tsukuba, Ibaraki 305-0047, Japan

Tel.: (+81)29-851-3354 ext. 2976

1. Introduction

Type II superconductors have been widely applied in strong magnets. NbTi with a superconducting transition temperature ($T_c$) of 10 K has been widely used as a superconducting magnet for nuclear magnetic resonance (NMR) spectrometers [1]. The resolution of the NMR spectrometer is strongly related to the intensity of the applied magnetic field. The maximum magnetic field of the NbTi magnet is limited to 10 T because of its upper critical field ($H_{c2}$ (0) ~ 20 T) [2, 3]. Therefore, superconductors with relatively high $H_{c2}$ are necessary for a high-field-magnet wire. In terms of these requirements, iron-based superconductors are one of the candidates due to high $H_{c2}$ over 100 T in maximum with its moderate anisotropy ($\gamma_{Hc2}$ 1 ~ 4) [4, 5].

FeSe which has the simplest crystal structure among iron-based superconductors, shows the $T_c$ around 8 K [6] and higher $H_{c2}$ (0) ~ 37.5 T [7] than NbTi. The high $H_{c2}$ nature with its low anisotropy and simplicity of the crystal structure make the FeSe superconductor a promising material for high-field-magnet wire applications. Therefore, a novel and easy method is required for the fabrication of superconducting tape or wire.

As a new way to synthesize FeSe superconductors, we have developed an electrochemical deposition technique [8–11], which can be deposited on various substrates by applying voltage to the substrates in the solution. This suggests the possible fabrication of FeSe superconducting tape by a reel-to-reel method: immersing a conducting tape into a solution and applying voltage. We have succeeded in observing the superconducting transition in such a tape by magnetization measurement [8-10]. However, zero resistivity has not been observed in as-electrodeposited FeSe samples. In this study, we attempted to achieve zero resistivity by improving the deposition process. It was found that the dipping time in a solution before applying voltage affected a surface state of the substrate and the compositional ratio of the samples. Here we report the first observation of zero resistivity in the as-electrodeposited FeSe film without any heat treatments.

## 2. Material and methods

The electrochemical deposition was carried out using a three-electrode method with Pt as a counter electrode and Ag/AgCl as a reference electrode. For a working electrode, on which FeSe is deposited, we used rolling-assisted biaxially textured substrate (RABiTS) tape. This tape has been widely used as a substrate in the production of $YBa_2Cu_3O_7$ coated conductors [12]. An aqueous solution was prepared by dissolving 0.03 M $FeCl_2·4H_2O$, 0.015 M $SeO_2$, and 0.1 M $Na_2SO_4$ in distilled water. All experiments were performed at a pH value of 2.1 and a solution temperature of 70°C. The application of voltage for the deposition of FeSe was carried out after dipping the tape into the solution for various dipping times (0, 5, 10, and 60 seconds). The voltage application time was fixed at 5 minutes in all experiments. Powder X-ray diffraction (XRD) was performed on the film deposited on RABiTS tape using a Mini Flex 600 (Rigaku) with Cu-K$α$ radiation. The surface morphology and chemical compositions were identified using a scanning electron microscope (SEM) equipped with an energy-dispersive X-ray (EDX) analyzer. The temperature dependence of resistivity was measured using a physical property measurement system (PPMS; Quantum Design) and a four-probe method down to 2.0 K. The Au electrodes were attached to the surface of the deposited film using a silver paste without peeling the film from the tape.

## 3. Results and discussion

In this paper, "dipping time" denotes the length of immersion in solution before applying voltage. Figure 1 shows the XRD patterns of the obtained films grown on the RABiTS tape with different applied voltages from -0.7 to -1.3 V for 5 min, where the dipping time was set at approximately 1 second. Only Se was deposited on the tape applied at -0.7 V. The single phase of tetragonal FeSe was obtained when applying voltage above -0.9 V. The high-quality crystalline FeSe films were obtained when applying voltages between -0.9 and -1.1 V. The XRD pattern for the film fabricated at -1.3 V indicates the phase formation of pure FeSe. This result indicates that an FeSe film can be produced on RABiTS tape by the electrochemical deposition technique in 5 minutes at a solution temperature of 70°C.

The compositional ratios of the obtained samples with various voltage applications are summarized in Fig. 2, where the dipping time before applying the voltage was set at approximately 1 second. With an increase in the negative bias voltage, the amount of Fe increased, whereas the amount of Se decreased. This result suggests that the compositional ratios of Fe and Se can be controlled by the bias voltage, which is consistent with previous reports [9]. At -1.1 V, the compositional ratio of Fe and Se was close to 1:1, indicating that the voltage of -1.1 V was optimal for forming FeSe. The thickness of the FeSe film synthesized at -1.1 V was measured from the cross-sectional SEM image and was about 10 µm.

We observed the surface of RABiTS tape with an SEM in order to investigate the influence on the surface state of the tape just before the FeSe deposition. The chemical composition of the surface of the tape, dipped in the solution without applying voltage, was analyzed using EDX spectroscopy. Figures 3 (a), (b) and (c) show the SEM image and EDX mapping result for Se and Fe of the surface of the tape above and in the solution after dipping in the solution for 10 seconds. Interestingly, the Se film was formed in the solution even though no voltage was applied. Moreover, the formation of the Se film was observed even after 1 second, and the amount of Se increased with extension of the dipping time. The formation of Se films is reasonable, because the metal substrate is immersed in the acid solution with a pH value of 2.1 at 70$^{o}$C, resulting in the reaction between the metal and the acid. This result indicates that FeSe is deposited on the thin Se film unless the dipping time is eliminated. The formation of the Se film changes the surface state of the tape before applying voltage and weakens the adhesion between the FeSe film and the tape.

Figure 4 summarizes the compositional ratios of Fe and Se for (a) the tape side and (b) the solution side, prepared at a bias voltage of -1.1 V for 5 minutes after various lengths of dipping time (0, 5, 10, and 60 seconds). A dipping time of 0 seconds was performed by applying voltage to the tape before inserting it into the solution. The EDX analysis for the tape side was carried out by peeling the FeSe film from the tape (see the schematic drawing in the insets), and no residue was observed on the tape after peeling the sample. This indicates that the Se film was peeled off with the FeSe film. The compositional ratios of Fe and Se were almost the same on both the tape side and the solution side at the dipping time of 0 seconds. On the other hand, the compositional ratios of Fe and Se on both sides varied from the ratio of Fe:Se = 1:1 with an extended dipping time. This result indicates that an increase in the dipping time triggers off-stoichiometry.

Figure 5 shows the temperature dependence of resistivity for the sample synthesized with various dipping times in the vicinity of the superconducting transition. In the sample fabricated with a dipping time of 0 seconds, the resistivity started to drop at $T_c^{onset}$ 8.4 K and reached zero at $T_c^{zero}$ 2.5 K. The $T_c^{onset}$ value is almost the same as the reported one [6, 7]. This is the first report on the observation of zero resistivity in an as-electrodeposited FeSe sample without any heat treatment. Although the $T_\mathrm{c}^\mathrm{onset}$ was also observed at 6.0 K for the sample with a dipping time of 5 seconds, zero resistivity was not observed until 2.0 K. With an increase in dipping time, the superconductivity began to be suppressed, and no superconducting transition was observed with a dipping time longer than 10 seconds. Furthermore, the resistivity in the normal state decreased with dipping time. These results can be explained by the deviation of the compositional ratio and the increase in the amount of Fe in the sample surface with the increase in dipping time, as shown in the EDX analysis. All of these results indicate that shortening the dipping time is the key factor in achieving zero resistivity in the as-electrodeposited FeSe.

## 4. Conclusion

We focused on the surface state of the substrate just before the electrochemical deposition and investigated the effect of dipping time before voltage application on the substrate and samples. Deviation of the compositional ratio of Fe:Se from 1:1 was found with increased dipping time. We successfully observed zero resistivity in the FeSe film by eliminating the dipping time. This is the first report on the achievement of zero resistivity in an as-electrodeposited FeSe sample without any heat treatment. These results suggest the possibility of low-cost fabrication of FeSe superconducting tape by the reel-to-reel method using the electrochemical deposition technique.

**Acknowledgments**

This work was partly supported by JST CREST, Japan, and JSPS KAKENHI Grant Number JP16J05432.

**Figure captions**

Fig. 1 XRD patterns of the obtained samples on the RABiTS tape with different voltage applications from -0.7 to -1.3 V. All of the indices indicate tetragonal FeSe. RABiTS tape and Se are shown as + and *, respectively.

Fig. 2 Compositional ratio of the obtained samples with various bias voltages.

Fig. 3 (a) SEM image and (b) EDX mapping result for Se and (c) Fe of tape above and in the solution after dipping in the solution for 10 seconds.

Fig. 4 Compositional ratios of Fe and Se for the (a) tape side and (b) solution side of the sample surface synthesized at a bias voltage of -1.1 V after various amounts of dipping time (0, 5, 10, and 60 seconds). "Dipping time" denotes the length of immersion in solution before applying voltage. Insets show the schematic drawing of the FeSe film on the RABiTS/Se film and the Se/FeSe film after peeling off for composition analyses.

Fig. 5 Temperature dependence of resistivity for the sample synthesized with dipping times of 0 and 5 seconds in the vicinity of the superconducting transition. (The inset shows temperature dependence of resistivity for the sample synthesized with dipping times of 10 and 60 seconds in the vicinity of the superconducting transition)

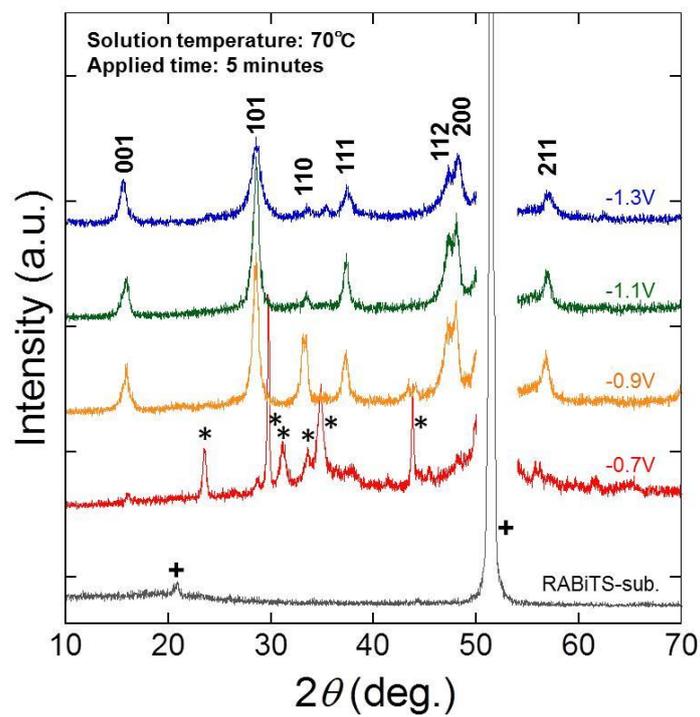

Fig. 1

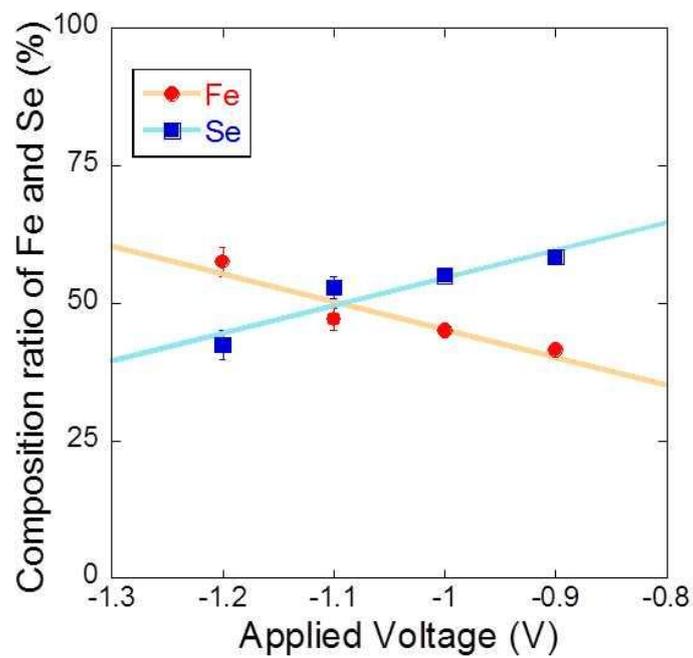

Fig. 2

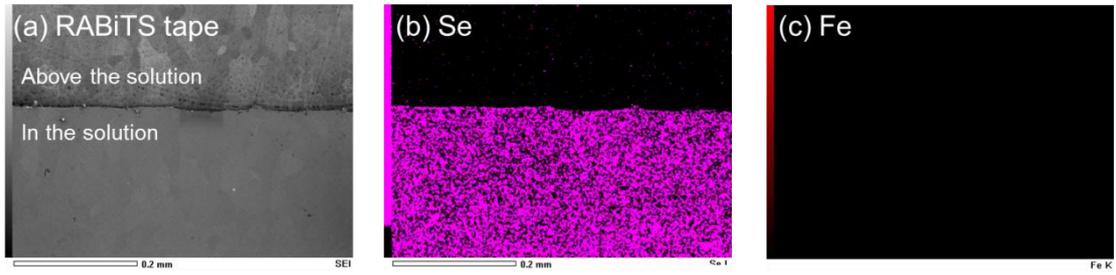

Fig. 3

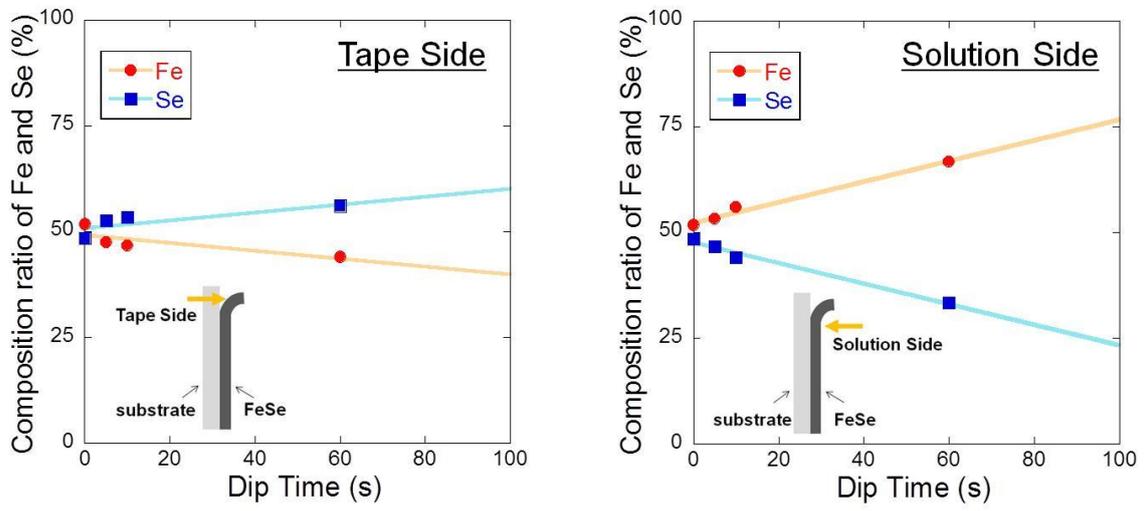

Fig. 4

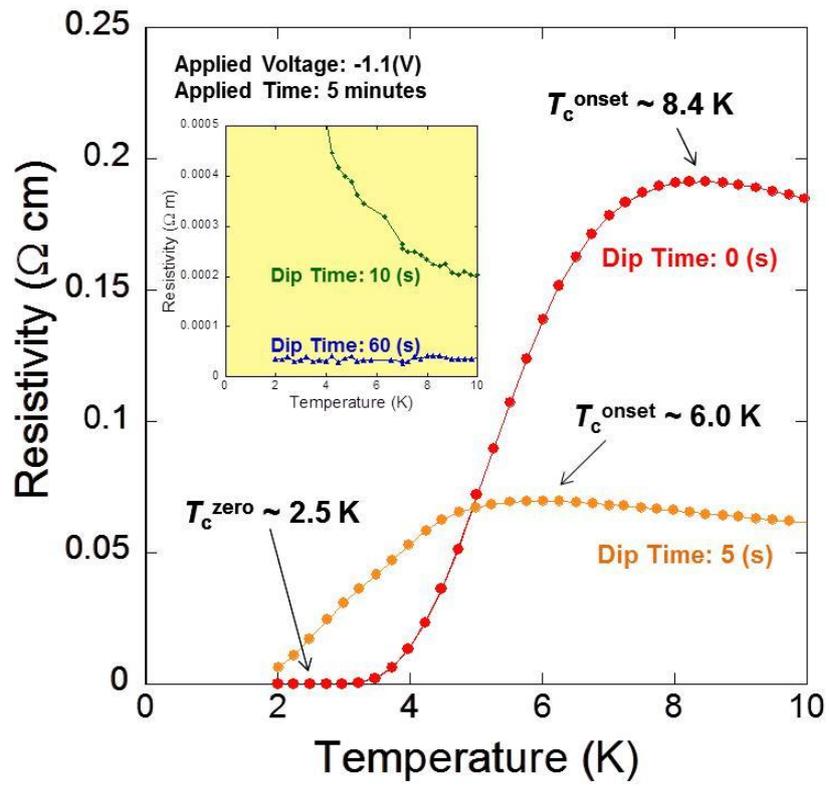

Fig. 5